\documentclass[aps,prb,twocolumn,superscriptaddress,showpacs]{revtex4}
\usepackage{graphicx}
\usepackage{amssymb}
\usepackage{bm}

\begin{document}


\title{Spin-related magnetoresistance of n-type ZnO:Al and Zn$_{1-x}$Mn$_{x}$O:Al thin films}


\author{T. Andrearczyk}
\email[]{andrea@ifpan.edu.pl}
\affiliation{Institute of Physics, Polish Academy of Science,
al.~Lotnik\'ow 32/46, 02-668 Warszawa, Poland}
\author{J. Jaroszy\'nski}
\altaffiliation[present address: ]{National High Magnetic Field Laboratory, Florida
State University, Tallahassee, Florida 32310}
\affiliation{Institute of Physics, Polish Academy of Science,
al.~Lotnik\'ow 32/46, 02-668 Warszawa, Poland}
\author{G. Grabecki}
\affiliation{Institute of Physics, Polish Academy of Science,
al.~Lotnik\'ow 32/46, 02-668 Warszawa, Poland}
\author{T. Dietl}
\affiliation{Institute of Physics, Polish Academy of Science,
al.~Lotnik\'ow 32/46, 02-668 Warszawa, Poland}
\affiliation{Institute of Theoretical Physics, Warsaw University,
00-681 Warszawa, Poland}
\affiliation{ERATO
Semiconductor Spintronics Project, Japan Science and Technology
Agency, al.~Lotnik\'ow 32/46, 02-668 Warszawa, Poland}
\author{T. Fukumura}
\affiliation{Institute for Materials Research, Tohoku University,
Sendai 980-8577, Japan}
\author{M. Kawasaki}
\affiliation{Institute for Materials Research, Tohoku University,
Sendai 980-8577, Japan}


\date{\today}

\begin{abstract}
Effects of spin-orbit coupling and s-d exchange interaction are
probed by magnetoresistance measurements carried out down to 50~mK
on ZnO and Zn$_{1-x}$Mn$_{x}$O with  $x = $ 3 and 7\%. The films
were obtained by laser ablation and doped with Al to electron
concentration  $\sim 10^{20}$~cm$^{-3}$. A quantitative
description of the data for ZnO:Al in terms of weak-localization
theory makes it possible to determine the coupling constant
$\lambda_{\mbox{\scriptsize so}}=(4.4\pm 0.4)\times 10^{-11}$~eVcm
of the $kp$ hamiltonian for the wurzite structure,
$H_{\mbox{\scriptsize so}} = \lambda_{\mbox{\scriptsize so}}
\bm{c}(\bm{s}\times \bm{k})$. A complex and large
magnetoresistance of Zn$_{1-x}$Mn$_{x}$O:Al is interpreted in
terms of the influence of the s-d spin-splitting and magnetic
polaron formation on the disorder-modified electron-electron
interactions. It is suggested that the proposed model explains the
origin of magnetoresistance observed recently in many magnetic
oxide systems.
\end{abstract}

\pacs{72.15.Rn,72.25.Rb,72.80.Ey,75.50.Pp}

\maketitle

In the emerging field of semiconductor
spintronics\cite{Wolf01Diet01Ohno02Zuti04} spin-orbit and sp-d
exchange interactions serve for spin manipulation in nonmagnetic
and magnetic semiconductors, respectively.  At the same time,
these interactions limit spin coherence time in these systems.
Within the standard Drude-Boltzmann theory of charge transport,
spin effects contribute only weakly to the conductivity. However,
electrons in doped wide band-gap semiconductors are at the
localization boundary, where charge transport is strongly affected
by quantum interference of both scattered waves and amplitudes of
the electron-electron interaction. Accordingly, the conductance in
these systems is sensitive to phase breaking mechanisms such as
spin relaxation as well as to symmetry lowering perturbations such
as the magnetic field and Zeeman splitting of electronic
states.\cite{Alts85Fuku85Lee85}

In this paper we compare magnetoresistance (MR) of non-magnetic
n-ZnO and paramagnetic n-(Zn,Mn)O. In both cases MR contains
positive and negative components but its overall shape and
magnitude is very different, a behavior reminiscent of that
observed previously for n-CdSe and n-(Cd,Mn)Se.\cite{Sawi86} A
quantitative description of the results demonstrates that MR of
ZnO is dominated by a destructive influence of the magnetic field
on interference of scattered waves. This orbital effect produces a
sizable negative MR, which below 1~mT is perturbed by the
so-called antilocalization caused by a spin-orbit coupling. We
determine the corresponding coupling constant and compare it to
results of available first principles computations. In the case of
paramagnetic n-(Zn,Mn)O, we show that MR  results primarily from
the influence of giant s-d spin splitting and spin scattering of
bound magnetic polarons on electron-electron interactions. These
effects produce large and strongly temperature dependent positive
and negative MR, respectively, the latter dominating in the
strongly localized regime, attained here due to strong self
compensation. We argue that our conclusions can be extended to
other oxides, providing information on magnetic order and on the
coupling of charge carriers to localized magnetic moments.

Our Zn$_{1-x}$Mn$_{x}$O thin films are grown by pulsed laser
deposition technique\cite{Fuku99} on sapphire (0001) substrates,
so that the c-axis of the wurzite structure is perpendicular to
the film plane. Technological Mn contents $x$ is 0, 0.03, and
0.07, for film thicknesses $d=440$, 310, and 310~nm, respectively.
These Mn concentrations are consistent with SIMS as well as with
the SQUID measurements.\cite{Sawi04} The SQUID data indicate that
antiferromagnetic interactions between the nearest neighbors are
relevant only. This causes a reduction of the paramagnetic spin
concentration to $x_{\mbox{\scriptsize eff}}=0.020$ and 0.038 for
the two (Zn,Mn)O samples in question if the Mn magnetic moment is
$5.0\mu_B$. An uniform Al doping with the concentration
$x_{\mbox{\scriptsize Al}} = 0.5$ and 4\% in the case of ZnO and
(Zn,Mn)O, respectively, results in the high and temperature
independent electron concentrations $n =1.8$, 1.4 and $1.1\cdot
10^{20}$~cm$^{-3}$, for $x=0$, 3, and 7\%, respectively, according
to Hall effect data. Transport measurements are performed in four
probe geometry in a $^3$He/$^4$He fridge down to 0.05~K, and in a
$^4$He cryostat above $T>1.5$~K by using a.c. lock-in technique.
Two types of electric contacts have been employed. Indium contacts
are easy to obtain by soldering method, but they obscure high
resolution measurements when indium becomes superconducting.
Accordingly, in the low temperature range we use gold spring
needles as contacts.

\begin{figure}[t]
\includegraphics{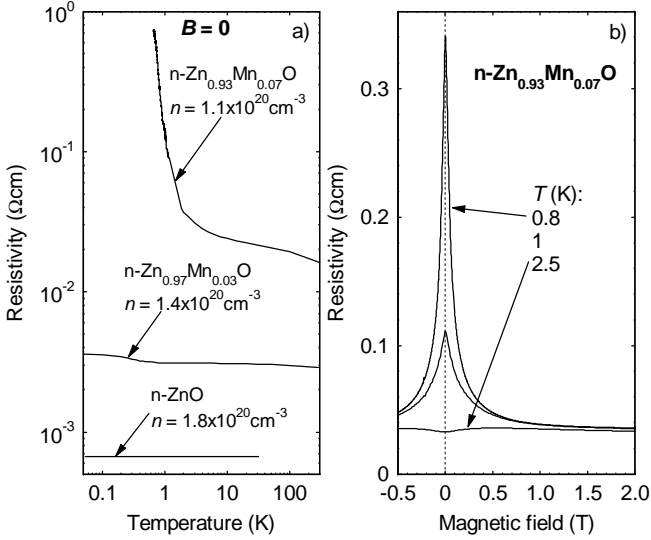}
\caption{Resistivity as a function of temperature
(a) and magnetic field (b) for  Zn$_{1-x}$Mn$_{x}$O and
Zn$_{0.93}$Mn$_{0.07}$O, respectively, revealing temperature dependent localization.}
\label{fig:rvst}
\end{figure}

Despite rather similar electron concentrations $n$, the
resistivities $\rho$  of particular samples are rather different,
as shown in Fig.~\ref{fig:rvst}(a). The determined values of the
diagonal and Hall resistivities lead to electron mobilities $\mu =
52$, 14, and 2~cm$^2$/Vs at 4.2 K for $x =0$, 3, and 7\%,
respectively. In order to elucidate the origin of such a strong
dependence $\mu(x)$ we first evaluate the effect of alloy and spin
disorder scattering assuming the band offset between the
conduction band of wurzite MnO and ZnO and the s-d exchange energy
to be $VN_{\mbox{\scriptsize o}} =2$~eV and $\alpha
N_{\mbox{\scriptsize o}} = 0.19$~eV, respectively, \cite{Diet94}
where  $N_{\mbox{\scriptsize o}} = 4.24\cdot 10^{22}$~cm$^{-3}$ is
the cation concentration in ZnO. For the effective mass $m^* =
0.3m_o$ the value of mobility limited by these scattering
mechanisms is then 690 and 330~cm$^2$/Vs  for $x = 3$ and 7\%,
respectively, much too high to explain the experimental values.

We note, however, that the large magnitude of
$VN_{\mbox{\scriptsize o}}$ may account for a growing role of
self-compensation with $x$, and elucidate why electrical activity
of Al donors decreases so strongly with the Mn incorporation.
Taking the static dielectric constant $\epsilon_s = 8$ and
assuming that the relevant acceptor defects are singly ionized,
that is that the total ionized impurity concentration is $N_{ii} =
2N_{\mbox{\scriptsize o}}x_{\mbox{\scriptsize Al}}-n$, we obtain
from the Brooks-Herring formula $\mu_{ii} = 82$, 5.0, and
4.1~cm$^2$/Vs for $x = 0$, 3, and 7\%, respectively. These values
correspond to $k_Fl= 16$, 0.86, and 0.6, pointing out that the
self-compensation drives the system towards Anderson-Mott
localization occurring when the product of the Fermi wave vector
$k_F$ and mean free path $l$ becomes smaller than
one.\cite{Alts85Fuku85Lee85} The proximity to the strongly
localized regime is confirmed by a substantial temperature
dependence of resistivity observed for the 7\% sample. The
resistivity increase is particularly strong below 1~K
[Fig.~\ref{fig:rvst}(a)], and signalizes the so-called temperature
dependent localization,\cite{Sawi86,Diet94,Ferr01} associated with
the formation of bound magnetic polarons. Colossal negative MR at
low temperatures, shown in Fig.~\ref{fig:rvst}(b), corroborates
this conclusion.

\begin{figure}
\includegraphics{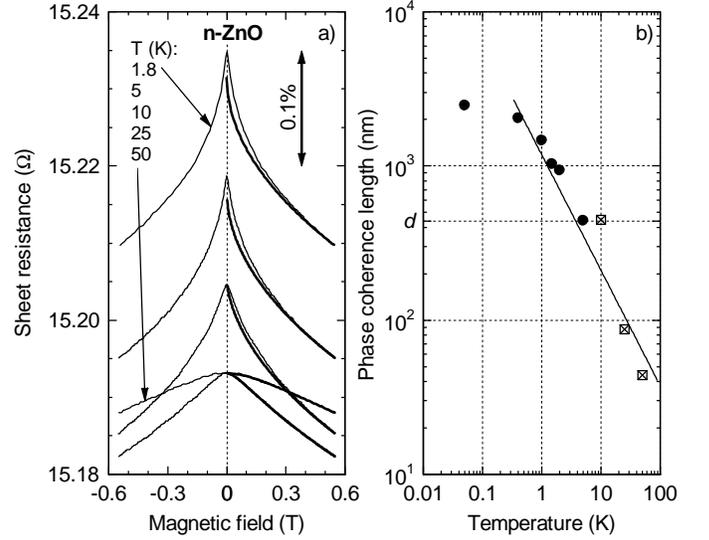}
\caption{Resistance changes in the magnetic field for n-ZnO (thin
lines) compared to calculations (thick lines) within the weak
localization theory for the 3D case. For $T=1.8$ and 5~K the phase
coherence length $L_{\varphi}$ is set to infinity, while the
fitted values of $L_{\varphi}$ at higher temperatures are plotted
in (b) by squares. Dots denote $L_{\varphi}$ determined by fitting
the weak-field data presented in Fig.~\ref{fig:zno0wf}. Straight
line shows the dependence $T^{-3/4}$ expected for $L_{\varphi}(T)$
in the 3D case.} \label{fig:zno0hf}
\end{figure}

Owing to the sufficiently large experimental values of $k_Fl$ in
ZnO and Zn$_{0.97}$Mn$_{0.03}$O, we expect MR in these samples to
be described by quantum corrections to conductivity in the weakly
localized regime.\cite{Alts85Fuku85Lee85,Sawi86}
Figure~\ref{fig:zno0hf}(a) depicts MR of ZnO in the field applied
perpendicularly to the film plane, that is along the c-axis. As
seen, MR is negative and temperature dependent, particularly in
weak magnetic fields. Such character and magnitude of MR, of the
order of 0.1\% at $B=0.5$~T, are similar to that observed for an
accumulation layer on ZnO.\cite{Gold99} We assign this MR to a
destructive effect of the magnetic field (vector potential) on
constructive interference corresponding to two time-reversal paths
along the same self-crossing trajectories. Thick lines in
Fig.~\ref{fig:zno0hf}(a) has been calculated within this model by
using Kawabata's three dimensional (3D)
formula,\cite{Alts85Fuku85Lee85,Kawa80} treating the phase
coherence length $L_{\varphi}(T)$ as a fitting parameter.  It is
apparent, however, that the model fails to describe the data
quantitatively at low temperatures and in low magnetic fields.  We
assign this failure to a dimensional cross-over, as according to
the values of $L_{\varphi}$ summarized in
Fig.~\ref{fig:zno0hf}(b), $L_{\varphi}$ becomes greater than the
sample thickness $d$ below 10~K.

\begin{figure}
\includegraphics{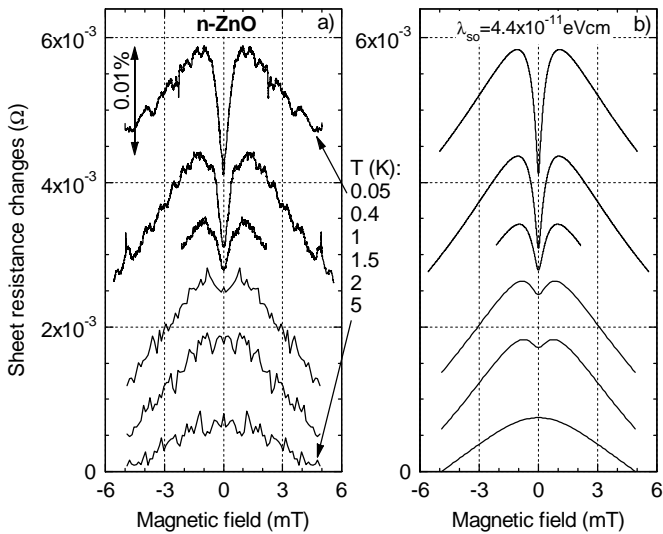}
\caption{Resistance changes in the magnetic field for n-ZnO (a)
compared to calculations (b) within the weak localization theory
for the 2D case. Curves are vertically shifted for clarity and
symmetrized.} \label{fig:zno0wf}
\end{figure}

Figure~\ref{fig:zno0wf} presents a comparison of experimental and
calculated MR of ZnO in the magnetic field below 6~mT. A
characteristic positive component of MR, signalizing the presence
of spin-orbit scattering, is detected below 1~mT at low
temperatures. Similarly to the case of n-CdSe,\cite{Sawi86} we
link this scattering to the presence of the term
$\lambda_{\mbox{\scriptsize so}} \bm{c}(\bm{s}\times \bm{k})$ in
the $kp$ hamiltonian of the wurzite structure, which in the
employed geometry leads to the spin-orbit relaxation rates
$\tau_{sox}^{-1}=\tau_{soy}^{-1}=\lambda_{\mbox{\scriptsize
so}}^2k_{F}^2\tau /12/\hbar^2$; $\tau_{soz}^{-1}=0$,\cite{Alts82}
where $\tau$ is momentum relaxation time. Since $L_{\varphi}(T) >
d$ and the magnetic length $L_B = (\hbar/eB)^{1/2} = d$ at 3.4~mT
we describe the data by Hikami {\em et al.} 2D
theory,\cite{Alts85Fuku85Lee85,Hika80} treating $L_{\varphi}(T)$
and $\lambda_{\mbox{\scriptsize so}}$ as the fitting parameters.
As shown in Fig.~\ref{fig:zno0wf}, we obtain a quite good
description of our findings with $\lambda_{\mbox{\scriptsize so}}
= 4.4\cdot 10^{-11}$~eVcm. Interestingly, the value of
$\lambda_{\mbox{\scriptsize so}}$, determined within the present
model with accuracy of about 10\%, is by a factor two greater than
that obtained by recent first principles computations.\cite{Lew96}
In contrast, the experimental value for CdSe,\cite{Sawi86,Dobr84}
$\lambda_{\mbox{\scriptsize so}} = (55 \pm 10)\cdot 10^{-11}$~eVcm
is by a factor of two smaller than the theoretical
result.\cite{Lew96} In any case, it is obvious that at a given
electron density a dramatic reduction of $\tau_{so}^{-1}$ occurs
on going from heavier CdSe to lighter ZnO. Finally, we note that
according to Fig.~\ref{fig:zno0hf}(b), $L_{\varphi}(T)\sim
T^{-3/4}$ down to ~0.5~K, as expected for the phase breaking by
electron-electron collisions in disordered 3D
systems.\cite{Alts85Fuku85Lee85} Since $L_T = (\hbar D/k_BT)^{1/2}
= d$ at 50~mK, we assign a weaker temperature dependence at low
temperatures to an onset of the dimensional cross-over, as in the
2D case $L_{\varphi}(T)\sim T^{-1/2}$.\cite{Alts85Fuku85Lee85}
However, at this stage, we cannot exclude perturbation effects of
stray fields or noise heating at the lowest temperatures, which
may affect the determined values of $L_{\varphi}(T)$.

\begin{figure*}
\includegraphics[width=7in]{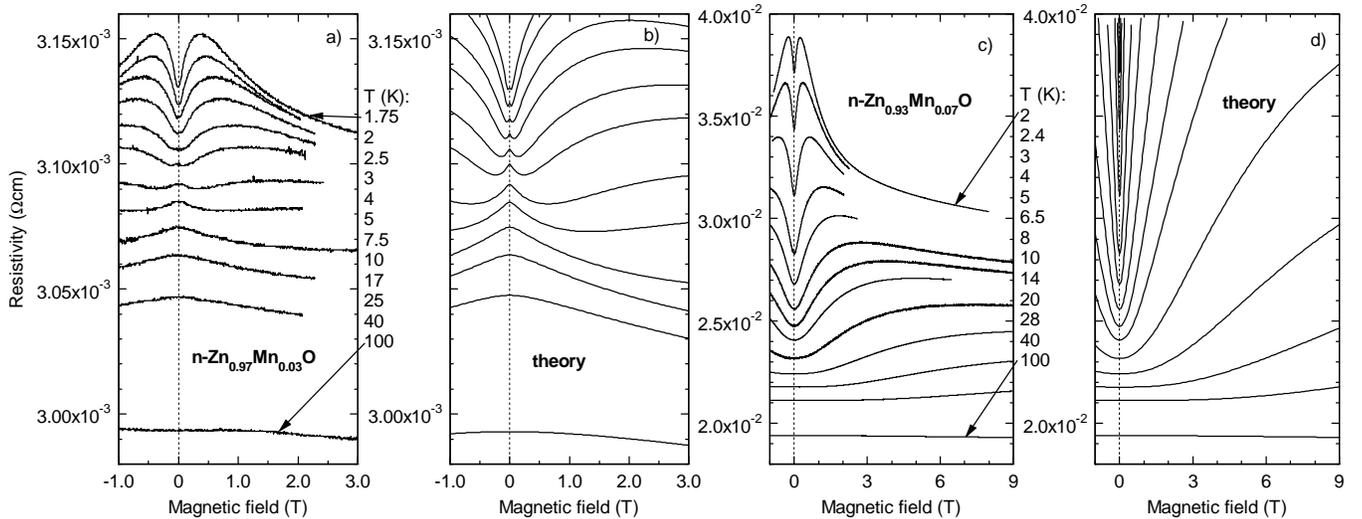}
\caption{Measured (a,c) and calculated with no adjustable
parameter (b,d) resistivity changes in the magnetic field for
n-Zn$_{0.97}$Mn$_{0.03}$O (a,b) and n-Zn$_{0.93}$Mn$_{0.07}$O
(c,d). Weak localization theory takes into account effects of the
magnetic field on interference of scattered waves and of the
spin-splitting on disorder-modified electron-electron interactions
but neglects effects associated with the formation of magnetic
polarons, which are thought to result in negative
magnetoresistance at low temperatures.} \label{fig:znmno}
\end{figure*}

We now turn to MR found for the films of paramagnetic (Zn,Mn)O,
shown in Fig.~\ref{fig:znmno}. A competition between positive and
negative contributions to MR is clearly visible. Despite that the
overall shape of MR is similar to that n-ZnO, its huge magnitude
as well as the field and temperature dependencies indicate that
effects brought about by the presence of the Mn spins dominate.
Actually, it has been shown
previously\cite{Sawi86,Diet94,Smor97,Andr02} that giant
spin-splitting of band states in diluted magnetic semiconductors
affects considerably quantum correction to the conductivity
associated with the disorder modified electron-electron
interactions.\cite{Alts85Fuku85Lee85} This results in positive
MR,\cite{Alts85Fuku85Lee85} if the Mn ions are not already
spin-polarized in the absence of the magnetic field. Furthermore,
the spin-splitting lead to a redistribution of the electrons
between spin subbands, which diminishes localization by increasing
the carrier kinetic energy.\cite{Fuku79} In our case, however, the
spin-splitting is more than ten times smaller than the Fermi
energy, which makes this mechanism of a minor importance. We
rather assign the negative component of MR to an enhancement of
spin-disorder scattering associated with the formation of bound
magnetic polarons on approaching the strongly localized
regime,\cite{Sawi86,Diet94} though quantitative theory of the
effect has not yet been elaborated. In line with this model,
negative MR becomes stronger with decreasing temperature and
increasing Mn content.

We execute  MR calculations for (Zn,Mn)O with no fitting parameter
taking into account both single-electron and many-body quantum
effects in the weakly localized regime, $k_{F}l\gg
1$.\cite{Alts85Fuku85Lee85}  Because of smaller magnitude of
diffusion constant in (Zn,Mn)O comparing to ZnO, we employ the 3D
formulae.\cite{Alts85Fuku85Lee85} The spin-splitting of the
conduction band contains the Zeeman term $g\mu_BH$, where $g =
2.0$ in ZnO, and the s-d contribution $x_{\mbox{\scriptsize eff}}
\alpha N_{\mbox{\scriptsize o}} SB_{S}(T,H)$, where the numerical
values of $x_{\mbox{\scriptsize eff}}$ and $\alpha
N_{\mbox{\scriptsize o}}$ have been given above and, in the
temperature range of interest here, $B_S$ is the paramagnetic
Brillouin function for $S = 5/2$. The parameter describing the
magnitude of electron-electron interaction in the triplet channel
is taken as $F_{\sigma} = 2g_3 =1$.  Furthermore, for
$\tau_{\varphi}(T)$ we adopt the values obtained for ZnO. As shown
in Figs.~\ref{fig:znmno}(a,b), the calculation for the 3\% sample
reproduces well the competition between negative  and positive
contributions to  MR at high temperatures. At lower temperatures,
however, only positive component is properly described in the
range of the weak fields. In high fields and at low temperatures,
in turn, we deal with an additional negative contribution, which
is too large to be explained by the employed weak-localization
theory. We regard this effect as a precursor of the polaron
formation. This situation is even more marked for $x=7$\% sample,
for which the comparison between measured and calculated MR is
depicted in Figs.~\ref{fig:znmno}(c,d). We note that
$k_{F}l\approx 2.4$ and $k_{F}l\approx 0.2$ for the $x=3$\% and
$x=7$\% samples, respectively, which means that the 7\% sample is
in the strongly localized regime, for which polaron effects should
be particularly important.

In summary, we have determined the magnitude of spin-orbit
relaxation time of electrons in ZnO by describing quantitatively
high-resolution low-temperature MR measurements in terms of
quantum corrections to the conductivity of disordered systems.
Within our interpretation, this spin scattering rate is
proportional to the square of the coupling constant between the
spin and momentum, in contrast to the corresponding band splitting
that is linear in $\lambda_{\mbox{\scriptsize so}}$. Accordingly,
wide-band gap semiconductors for which $\lambda_{\mbox{\scriptsize
so}}$ is small, appear to be particularly suitable for spin
manipulation by spin-orbit effects. A small magnitude of
$\lambda_{\mbox{\scriptsize so}}$ makes also the corresponding MR
to become dominated, in rather weak magnetic fields, by MR
associated with the influence of the vector potential on
interference of self-crossing trajectories. Interestingly, this
orbital effect makes {\em negative} MR to be a characteristic
feature of {\em non-magnetic} semiconductors, while the presence
of a giant spin-splitting, specific to diluted magnetic
semiconductors in a paramagnetic phase, gives rise to large
positive MR. This is in contrast to both diluted magnetic metals
and magnetic semiconductors in the strongly localized regime, in
which negative, not positive, MR points to the presence of a
coupling between localized spins and charge carriers. In diluted
magnetic semiconductors at the localization boundary, there is a
coexistence of positive and negative MR caused by the
spin-splitting and the formation of bound magnetic polarons,
respectively. The latter takes over at low temperatures and to our
knowledge has not yet been described theoretically. Importantly,
the model of MR invoked here appears to explain  qualitatively the
findings reported previously for (Zn,Mn)O,\cite{Fuku99}
(Zn,Co)O,\cite{Kim03} (Zn,Fe)O,\cite{Han02}
(Sn,Mn)O$_{2}$,\cite{Kimu02} (Ti,Co)O$_{2}$,\cite{Shin03} to quote
only a few examples. In particular, the presence of sizable
positive MR, suggests that the band splitting vanishes in the
absence of an external magnetic field.

We thank Hideo Ohno and Maciej Sawicki for valuable discussions.
This work was partly supported by PBZ/KBN/044/P03/2001 Grant
and ERATO Semiconductor Spintronics Project.

\end{document}